\documentclass{ifacconf}
\usepackage{natbib}
\usepackage[english]{babel}
\usepackage{mathtools}
\usepackage{amsmath,amsfonts}
\usepackage{ragged2e}
\usepackage{graphicx}
\usepackage{tikz}
\usepackage{pgfplots} 
\usepackage{pgfgantt}
\usepackage{pdflscape}
\pgfplotsset{compat=newest} 
\pgfplotsset{plot coordinates/math parser=false}
\usepackage{caption}
\usepackage{subcaption}
\usepackage{ulem}
\usepackage{color}
\usepackage{booktabs,ragged2e}
\usepackage[flushleft]{threeparttable}
\usetikzlibrary{backgrounds}
\usetikzlibrary{calc}
\usepackage{gensymb}
\usepackage{romannum}
\usetikzlibrary{intersections, calc, angles}
\usetikzlibrary{arrows,decorations.markings}
\begin{document}

\begin{frontmatter}
\title{Cooperative Control in Eco-Driving of Electric Connected and Autonomous  Vehicles in an Un-Signalized Urban Intersection} 

\author[First]{Vinith Kumar Lakshmanan}
\author[Second]{Antonio Sciarretta}
\author[Second]{Ouafae El Ganaoui-Mourlan}
\date{March 2022}

\address[First]{IFP Energies Nouvelles, Rueil-Malmaison, France (e-mail: name-middlename.surname@ifpen.fr)}
\address[Second]{IFP Energies Nouvelles, Rueil-Malmaison, France (e-mail: name.surname@ifpen.fr)}

\begin{abstract}    
This paper addresses the problem of finding the optimal Eco-Driving (ED) speed profile of an electric connected and automated vehicle in an isolated urban un-signalized intersection. The problem is solved using Pontryagin's Minimum Principle (PMP) and analytical solutions are presented for various conflicts that occur in an intersection. Cooperation is also introduced amongst the CAVs as the ability to share intentions. Two levels of cooperation, namely the cooperative and non-cooperative ED algorithms are evaluated, in a simulation environment, for energy efficiency with Intelligent Driver Model as baseline.  
\end{abstract}

\begin{keyword}
Eco-driving, cooperation, nonlinear and optimal control, intersection
\end{keyword}

\end{frontmatter}

\section{Introduction}
The main goal of Eco-Driving (ED) is to adopt an energy-efficient driving technique and people motivated by this technique are often called hypermilers. In the past decade, ED has been formulated as an optimal control problem, where the vehicle speed is directly controlled or indirectly advised to the driver to minimize energy consumption over a certain horizon (\cite{7587980,WANG2014271,8286942,6043133}). ED can be applied to several driving scenarios such as car-following, intersection crossing, eco-acceleration/deceleration, etc.  

With the advent of Connected and Automated Vehicles (CAVs), ED can be enforced more easily than with human drivers.  Furthermore, in a connected environment, the CAVs can communicate to cooperate amongst each other rather than compete against each other. Cooperation in what follows refers to the sharing of information and the willingness of the CAVs to coordinate their movement for the common good. Based on the amount of information shared and the coordination of movements, cooperation can be categorized into three types, i.e., non-cooperative, cooperative, and centralized cooperative schemes. In the non-cooperative scheme, each CAV selfishly optimizes for itself and shares only its instantaneous control action with its neighbors. In the cooperative scheme, each CAV still selfishly optimizes for itself but shares its future intentions with the neighboring CAVs and in the centralized cooperative scheme, the control action of each CAV is such that it optimizes for the entire group. Control schemes of single-CAV 
optimization (i.e., non-cooperative and cooperative), often fall under the umbrella of decentralized optimization in literature. 
In this paper, two types of cooperation, namely the \textit{non-cooperative and cooperative} schemes are applied to the \textit{intersection crossing} ED scenario. 

An intersection is a junction where two or more roads meet and are categorized into two-way, three-way, four-way intersections, roundabouts, etc. They can also be categorized based on traffic control technology as: signalized (traffic lights), stop, yield and un-signalized intersections. An intersection being a shared resource and CAVs wanting to use it simultaneously can cause conflicts leading to either rear-end or lateral collision. In this paper, we consider a single lane, isolated four-way un-signalized intersection. In the literature on un-signalized intersections, the area within the perimeter where CAVs can communicate with each other, or infrastructure is often called the Control Zone (CZ) and the region in the center, where vehicle paths cross, is called the Intersection Zone (IZ). The lane in the CZ leading to and out of the IZ is called the entry lane and exit lane, respectively. With an intersection being a shared resource it poses two main challenges, namely, scheduling of the CAVs in the IZ and the continuous optimization, involving the motion planning of CAVs.

The scheduling problem determines the priority of CAVs at the intersection. Heuristic or rule-based methods such as First Come First Serve (FCFS) (\cite{DresnerS08}) or First-In-First-Out (FIFO), right-before-left and nearest to the crossing point, etc. are often used in literature to schedule CAVs crossing an intersection. Scheduling problem can also be cast as an optimization problem with the objective to minimize travel time or energy consumption using tools such as Mixed Integer Linear Programming (MILP) (\cite{milp1,milp2}). The second challenge, motion planning, involves generating paths or trajectories for the longitudinal and lateral motion of the CAVs. The literature distinguishes path as having a spatial component and trajectory having a temporal component. A path is geometric set of points, $f(x,y,,..)=0$, to go from point \textit{a} to point \textit{b} and trajectory describes the evolution of path in time, $s(t)$. 
With a CAV following a predefined fixed path, the problem narrows down to finding the trajectory (i.e., velocity) with respect to a certain objective while respecting constraints such as vehicle dynamics, speed limits, and safety constraints. In our work, we considered a predefined schedule/crossing order and a fixed path.  

A rich body of literature is available for intersection scheduling  and motion planning. A comprehensive overview of the various heuristic and optimization methods employed can be found in (\cite{revMal,revChen}). With a focus on trajectory optimization with respect to energy consumption, we review some of the works here.
(\cite{Bichiou2019}) proposes a problem formulation to minimize the trip time and the control effort of a CAV crossing an intersection.  Pontryagin’s Minimum Principle (PMP) is used to formulate the problem. The authors state that the problem, in theory, could be solved using PMP. However, owing to the difficulty in obtaining analytical solutions, the problem is solved using numerical discrete and convex optimization. Rear-end collision in the CZ is avoided by formulating a position-inequality constraint and lateral collision is avoided by modeling the entire IZ as a collision region.  Collision region can also be modelled as a point on the fixed path of the vehicles allowing for more than one vehicle inside the IZ. The  red dots in Fig.~\ref{fig:CC_points} represent a collision point. In the numerical Model Predictive Control (MPC) proposed by (\cite{hult}), the cost functional directly captures the energy usage of an Electric Vehicle (EV) and the travel time. 
(\cite{makarem,campos,bassam}) formulate a quadratic MPC seeking to minimize the control effort and the deviation from the reference velocity. 

 While numerical methods facilitate the use of  non-linear models and complex formulations, analytical methods provide fast and explicable solutions. However, only a handful of research efforts have been made on obtaining analytical closed-form solutions in the optimal control framework of CAVs in an intersection. (\cite{MALIKOPOULOS2018244}) presents a bi-level optimization problem for scheduling and trajectory of CAVs in an intersection without any turns. The upper-level optimization schedules the CAVs by maximizing the throughput 
 under the FIFO policy.  The lower-level problem minimizes acceleration, which the authors argue minimizes the transient engine operation and in turn the fuel consumption. With the optimization horizon of the lower-level problem only on the entry lane, the solution to this problem produces an optimal speed profile only until the start of IZ. The vehicles are restricted to a constant velocity thereafter through the intersection.   In a follow-up work (\cite{ZHANG2019108563}), the authors extended the problem formulation to consider left and right turning vehicles. 
 The authors present a bi-level optimization solely for the trajectory of the CAVs. Still, under FIFO policy, the upper-level problem has an optimization horizon only for the entry lane jointly minimizing travel time and accelerations. The arrival time and velocity at the end of the entry lane are then used as input to the second optimization problem. The lower-level problem jointly reducing jerks and accelerations with an optimization horizon only for the IZ. The CAVs are restricted to constant velocity on the exit lane to avoid rear-end collision. The solution to the above formulation leads to one optimal speed profile on the entry lane and a second optimal speed profile inside the IZ followed by a constant velocity. 

The main objective of the paper is to obtain analytical solutions to the ED problem of an electric CAV crossing an un-signalized intersection subject to safety constraints, and to explore the benefits of cooperation . The main contributions/novelty of the paper can be summarised as follows.
We formulate a single-level optimization problem to the ED intersection scenario with an optimization horizon that includes the entry lane, the IZ and the exit lane. 
The analytical objective function used captures directly the energy usage in an EV, and can deal with position and speed constraints, similar to what has already been studied (\cite{9564680,LAKSHMANAN2021133,Han.2019}) and experimentally demonstrated (\cite{edexp}) for car-following.
Cooperation is also introduced amongst the CAVs as the ability to share intentions and two levels of cooperation, namely the cooperative and non-cooperative ED algorithms, are evaluated for performance in terms of energy consumption.

The organization of the paper is as follows. In section 2, the intersection and vehicle model along with its assumptions are presented. The various conflicts in an intersection are also discussed. In section 3, the optimal control problem is formulated along with the various constraints and its solutions are presented. Section 4, details the algorithm and the cooperative nature of the vehicles. The simulation results and its discussions are presented in section 5.

\section{Intersection Model, Vehicle Model and Conflicts}

This section describes the considered intersection and vehicle model along with its various assumptions. The various conflicts that occur in an intersection is also discussed.

\subsection{Intersection Model}
\begin{figure}
    \begin{subfigure}[b]{\columnwidth}
        \centering
        \begin{tikzpicture}[scale = 0.5,every node/.style={scale=0.8}]
\draw[black,very thick](3,0)--(3,2.5);
\draw[black,very thick](2.5,3) -- (0,3);
\draw[black,very thick](2.5,3) .. controls(2.85,2.85) .. (3,2.5);
\draw[black,very thick](5,0)--(5,2.5);
\draw[black,very thick](5.5,3) -- (8,3);
\draw[black,very thick](5.5,3) .. controls(5.15,2.85) .. (5,2.5);
\draw[black,very thick](3,8)-- (3,5.5);
\draw[black,very thick](2.5,5)--(0,5);
\draw[black,very thick](3,5.5) .. controls(2.85,5.15) .. (2.5,5);
\draw[black,very thick](5,8)--(5,5.5);
\draw[black,very thick](5.5,5)--(8,5);
\draw[black,very thick](5,5.5) .. controls(5.15,5.15) .. (5.5,5);
\draw[lightgray,dashed,thin](0,4)--(8,4);
\draw[lightgray,dashed,thin](4,0)--(4,8);

\draw[gray,dashdotdotted, thin](2.5,3) -- (2.5,5);
\draw[gray,dashdotdotted, thin](3,2.5) -- (5,2.5);
\draw[gray,dashdotdotted, thin](3,5.5) -- (5,5.5);
\draw[gray,dashdotdotted, thin](5.5,3) -- (5.5,5);

\draw[lightgray, ultra thin, <->](5.5,2.5) -- (8,2.5);
\node at (6.5,2.5) [below]{$\ell$};
\draw[lightgray, ultra thin, <->](3,-0.25) -- (4,-0.25);
\node at (3.5,-0.25) [below] {\it{w}};
\filldraw[black] (3.78,3.5) circle (0.05) node[above]{$\mathcal{C}$};
\filldraw[black] (2.5,3.5) circle (0.05) node[above]{$\mathcal{D}$};
\filldraw[black] (3.5,2.5) circle (0.05) node[below]{$\mathcal{M}$};
\draw[->,>=stealth,color=red] (0,3.5)-- (8,3.5); 
\node at (0,3.5)[left, color=red]{$p$};
\node at (8,4.5)[right, color=blue]{$i$};
\draw[->,>=stealth,color=blue](8,4.5)--  (5.5,4.5) arc[start angle=90, end angle=180,radius=2] -- (3.5,2.5) -- (3.5,0);

\node at (4,5) [above] {IZ};
\node at (2,7) [below] {CZ};

\end{tikzpicture}
        \caption{Intersection Model : $\mathcal{M},\mathcal{D},\mathcal{C}$ represent the merging point, diverging point and crossing point respectively. $\ell$ and $\it{w}$ are the exit and entry lane length and width of each lane respectively}
         \label{fig :intModel}
    \end{subfigure}
    \begin{subfigure}[b]{\columnwidth}
        \centering
        \includegraphics[scale = 0.4]{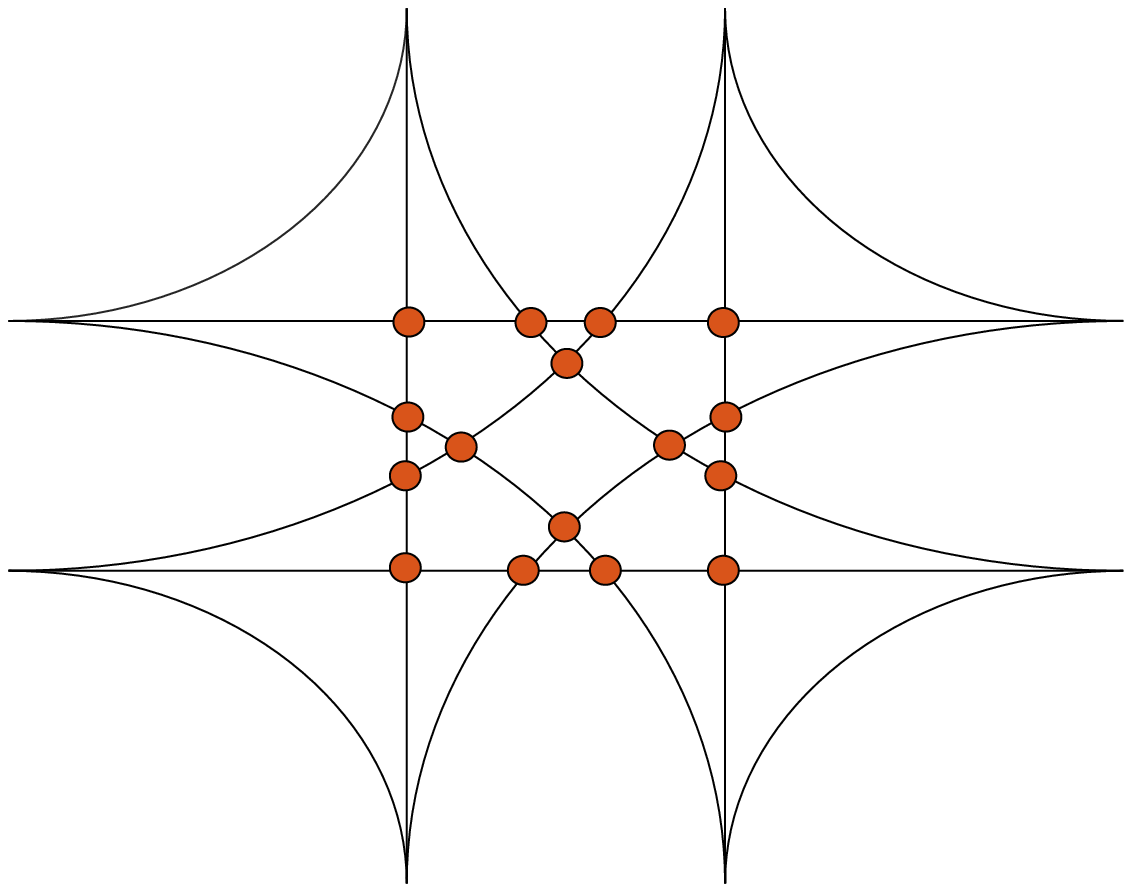}     \caption{The dots represents the 16 crossing conflict points}
        \label{fig:CC_points}
    \end{subfigure}
    \caption{Intersection Model and Conflict Model}
\label{fig:int model}
\end{figure}

The intersection considered here is an isolated un-signalized four legged intersection with flat straight roads crossing at right angles to each other. Each leg of the intersection consists of two lanes with traffic flowing in opposite directions. The center of the intersection where where two or more paths can intersect, i.e., cause a lateral collision is called the Intersection Zone (IZ). The roads leading to and away from the IZ, of length \textit{l}, are called the entry lane and exit lane respectively. The area around the IZ where CAVs can communicate with each other and a coordinator is called the Control Zone (CZ). The point where entry lane meets the IZ is called the Diverging Point ($\mathcal{D}$) and the point where IZ meets the exit lane is called the Merging Point ($\mathcal{M}$). Fig.~\ref{fig:int model} represents the intersection model considered in this paper.

For a four-way intersection with one lane in each direction, there are a total of 12 paths across the intersection. The entry lane and the direction of a CAV determines along which of the 12 paths it travels. Collision region in this work is modelled as points, called Crossing Points ($\mathcal{C}$), where the paths of the CAVs intersect. Figure~\ref{fig:CC_points} shows the the paths and $\mathcal{C}$ in the IZ.

We assume that each CAV follows a predefined path without any deviation. The path of a CAV taking a left or right turn is modelled as constant radius arc, with a central angle of $90\degree$, when passing through the IZ.

Let $\mathcal{N}(t) = \{1,2,..., N(t)\}$ be the set of CAVs' ID inside the CZ at time $t \in \mathbb R^+$, where $N(t) \in \mathbb{N}$ is  total number of CAV present in the CZ at $t$.  $\mathcal{CO}(t)$ is a permutation of $\mathcal{N}(t)$, that represents the crossing order, according to a given criterion. As mentioned earlier, the criterion can be FCFS, right-before-left, or as a result of an upper-level optimizer. In this work, we assume a predefined crossing order that is computed and communicated to the vehicles via a coordinator present at the intersection. 

Each CAV entering the intersection has perfect information about the geometry of the intersection and can compute the distance to the crossing points on its path. Furthermore, each CAV is able to receive information from other CAVs in the CZ, that have higher priority, on their entry lane, heading direction, arrival time at their crossing point, and instantaneous or future control actions. 
In this modelling framework, we impose further assumptions to abstract from implementation issues and focus on the fundamental aspect of motion planning: (\romannum{1}) all CAVs are equipped with  V2V and V2X communication capabilities and appropriate sensors to sense local information of itself and others in the proximity without losses or delays; (\romannum{2}) the CAVs are not allowed to overtake each other nor make a U turn; (\romannum{3}) no pedestrians are considered crossing the intersection.

\subsection{Vehicle Model}
As mentioned above, the CAVs follow a fixed path meaning vehicles can control only its acceleration/deceleration. This enables to decouple path-trajectory and use a second order longitudinal dynamics to describe each CAV $i \in \mathcal{N}(t)$.
Let $v_i(t)$, $x_i(t)$, $a_i$, $l_i$ denote the speed, position, acceleration and the length of the $i$-th  vehicle.
 With the control variable $u_i$ of each CAV chosen as the net force produced by the powertrain per unit mass, $F_t/m$, and the resistive forces represented by a constant $h$, the linearised longitudinal vehicle model is given as
\begin{equation}\label{vehicle_model}
\begin{aligned}
  \dot{x}_i &= v_i(t), \\
  \dot{v}_i &= u_i(t) - h.
\end{aligned}
\end{equation}

The energy consumption at the battery $P_b$ is calculated using simple transmission and electric motor models, inspired by those used in (\cite{6043133}) given by,
\begin{equation}\label{Pb}
    P_b(t) = p_0u_i(t)v_i(t) + p_1u_{i}^2(t),
\end{equation}
where $p_0$ and $p_1$ represent battery modelling parameters. 

\begin{figure*}[t!]
     \centering
     \begin{subfigure}[t]{0.2\textwidth}
         \centering
         \begin{tikzpicture}[scale = 0.4,every node/.style={scale=0.8}]
\draw[black,very thick](3,0)--(3,2.5);
\draw[black,very thick](2.5,3) -- (0,3);
\draw[black,very thick](2.5,3) .. controls(2.85,2.85) .. (3,2.5);
\draw[black,very thick](5,0)--(5,2.5);
\draw[black,very thick](5.5,3) -- (8,3);
\draw[black,very thick](5.5,3) .. controls(5.15,2.85) .. (5,2.5);
\draw[black,very thick](3,8)-- (3,5.5);
\draw[black,very thick](2.5,5)--(0,5);
\draw[black,very thick](3,5.5) .. controls(2.85,5.15) .. (2.5,5);
\draw[black,very thick](5,8)--(5,5.5);
\draw[black,very thick](5.5,5)--(8,5);
\draw[black,very thick](5,5.5) .. controls(5.15,5.15) .. (5.5,5);
\draw[lightgray,dashed,thin](0,4)--(8,4);
\draw[lightgray,dashed,thin](4,0)--(4,8);
\draw[gray,dashdotdotted, thin](2.5,3) -- (2.5,5);
\draw[gray,dashdotdotted, thin](3,2.5) -- (5,2.5);
\draw[gray,dashdotdotted, thin](3,5.5) -- (5,5.5);
\draw[gray,dashdotdotted, thin](5.5,3) -- (5.5,5);
\draw[->,>=stealth,color=red](1,3.5)--  (2.5,3.5) arc[start angle=270, end angle=360,radius=2] -- (4.5,5.5) -- (4.5,8) node at (4.5,8)[left, color=red]{\textit{d}};
\draw[->,>=stealth,color=blue,thick](0,3.5)--(7,3.5) node at (0,3.5)[above, color=blue]{\textit{i}};
\end{tikzpicture}
         \caption{Diverging Conflict}
         \label{fig:DP conflict}
     \end{subfigure}
     ~\hfill
     \begin{subfigure}[t]{0.2\textwidth}
         \centering
         \begin{tikzpicture}[scale = 0.4,every node/.style={scale=0.8}]
\draw[black,very thick](3,0)--(3,2.5);
\draw[black,very thick](2.5,3) -- (0,3);
\draw[black,very thick](2.5,3) .. controls(2.85,2.85) .. (3,2.5);
\draw[black,very thick](5,0)--(5,2.5);
\draw[black,very thick](5.5,3) -- (8,3);
\draw[black,very thick](5.5,3) .. controls(5.15,2.85) .. (5,2.5);
\draw[black,very thick](3,8)-- (3,5.5);
\draw[black,very thick](2.5,5)--(0,5);
\draw[black,very thick](3,5.5) .. controls(2.85,5.15) .. (2.5,5);
\draw[black,very thick](5,8)--(5,5.5);
\draw[black,very thick](5.5,5)--(8,5);
\draw[black,very thick](5,5.5) .. controls(5.15,5.15) .. (5.5,5);
\draw[lightgray,dashed,thin](0,4)--(8,4);
\draw[lightgray,dashed,thin](4,0)--(4,8);
\draw[gray,dashdotdotted, thin](2.5,3) -- (2.5,5);
\draw[gray,dashdotdotted, thin](3,2.5) -- (5,2.5);
\draw[gray,dashdotdotted, thin](3,5.5) -- (5,5.5);
\draw[gray,dashdotdotted, thin](5.5,3) -- (5.5,5);
\draw[->,>=stealth,color=blue](0,3.5)--  (2.5,3.5) arc[start angle=270, end angle=360,radius=2] -- (4.5,5.5) -- (4.5,8) node at (4.5,8)[left, color=blue]{\textit{i}};
\draw[->,>=stealth,color=red, thick] (4.5,0) -- (4.5,2.5) arc[start angle=0, end angle=90,radius=2] -- (2.5,4.5) -- (0,4.5) node at (4.5,0)[left, color=red]{\textit{c}};
\end{tikzpicture}
         \caption{Crossing Conflict}
         \label{fig:Crossing conflict}
     \end{subfigure}
     ~\hfill
     \begin{subfigure}[t]{0.2\textwidth}
         \centering
         \begin{tikzpicture}[scale = 0.4,every node/.style={scale=0.8}]
\draw[black,very thick](3,0)--(3,2.5);
\draw[black,very thick](2.5,3) -- (0,3);
\draw[black,very thick](2.5,3) .. controls(2.85,2.85) .. (3,2.5);
\draw[black,very thick](5,0)--(5,2.5);
\draw[black,very thick](5.5,3) -- (8,3);
\draw[black,very thick](5.5,3) .. controls(5.15,2.85) .. (5,2.5);
\draw[black,very thick](3,8)-- (3,5.5);
\draw[black,very thick](2.5,5)--(0,5);
\draw[black,very thick](3,5.5) .. controls(2.85,5.15) .. (2.5,5);
\draw[black,very thick](5,8)--(5,5.5);
\draw[black,very thick](5.5,5)--(8,5);
\draw[black,very thick](5,5.5) .. controls(5.15,5.15) .. (5.5,5);
\draw[lightgray,dashed,thin](0,4)--(8,4);
\draw[lightgray,dashed,thin](4,0)--(4,8);
\draw[gray,dashdotdotted, thin](2.5,3) -- (2.5,5);
\draw[gray,dashdotdotted, thin](3,2.5) -- (5,2.5);
\draw[gray,dashdotdotted, thin](3,5.5) -- (5,5.5);
\draw[gray,dashdotdotted, thin](5.5,3) -- (5.5,5);
\draw[->,>=stealth,color=red, thick] (4.5,0) -- (4.5,2.5) arc[start angle=180, end angle=90,radius=1] -- (5.5,3.5) -- (8,3.5) node at (4.5,0)[left, color=red]{\textit{e}};
\draw[->,>=stealth,color=blue,thick](0,3.5)--(5.5,3.5) node at (0,3.5)[below, color=blue]{\textit{i}};
\end{tikzpicture}
         \caption{Merging Conflict}
         \label{fig:MP conflict}
     \end{subfigure}
     ~\hfill
     \begin{subfigure}[t]{0.2\textwidth}
         \centering
         \begin{tikzpicture}[scale = 0.4,every node/.style={scale=0.8}]
\draw[black,very thick](3,0)--(3,2.5);
\draw[black,very thick](2.5,3) -- (0,3);
\draw[black,very thick](2.5,3) .. controls(2.85,2.85) .. (3,2.5);
\draw[black,very thick](5,0)--(5,2.5);
\draw[black,very thick](5.5,3) -- (8,3);
\draw[black,very thick](5.5,3) .. controls(5.15,2.85) .. (5,2.5);
\draw[black,very thick](3,8)-- (3,5.5);
\draw[black,very thick](2.5,5)--(0,5);
\draw[black,very thick](3,5.5) .. controls(2.85,5.15) .. (2.5,5);
\draw[black,very thick](5,8)--(5,5.5);
\draw[black,very thick](5.5,5)--(8,5);
\draw[black,very thick](5,5.5) .. controls(5.15,5.15) .. (5.5,5);
\draw[lightgray,dashed,thin](0,4)--(8,4);
\draw[lightgray,dashed,thin](4,0)--(4,8);
\draw[gray,dashdotdotted, thin](2.5,3) -- (2.5,5);
\draw[gray,dashdotdotted, thin](3,2.5) -- (5,2.5);
\draw[gray,dashdotdotted, thin](3,5.5) -- (5,5.5);
\draw[gray,dashdotdotted, thin](5.5,3) -- (5.5,5);
\draw[->,>=stealth,color=red, thick] (3,3.5)-- (8,3.5) node at (3,3.5)[below, color=red]{\textit{p}};
\draw[->,>=stealth,color=blue,thick](0,3.5)--(2.9,3.5) node at (0,3.5)[above, color=blue]{\textit{i}};
\end{tikzpicture}
         \caption{Car-Following Conflict}
         \label{fig:CF }
     \end{subfigure}
        \caption{Four Basic Constraints in an Intersection. }
        \label{fig:constraints}
\end{figure*}
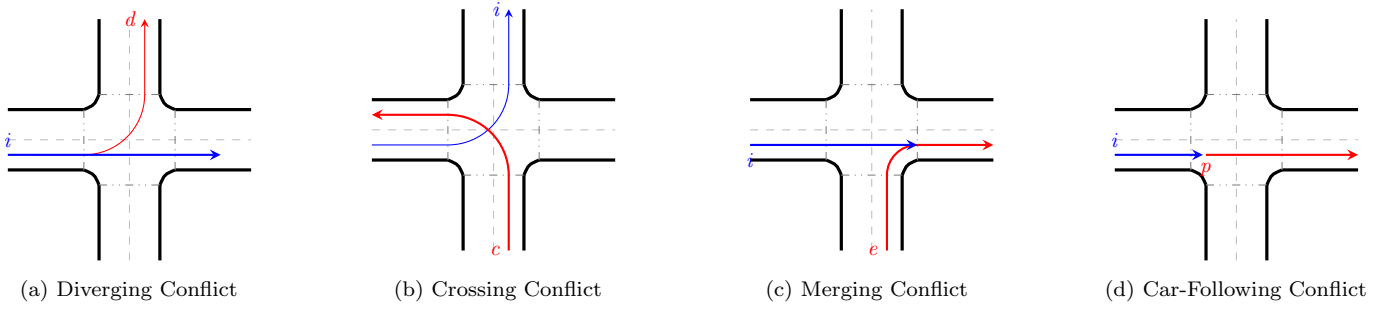

 \subsection{Conflicts}
There are four types of basic conflicts in a traffic conflict analysis: diverging, crossing, merging, and sequential/car-following conflict (\cite{conflict}). See Fig.~\ref{fig:constraints}. 

A diverging conflict occurs when CAVs have the same entry lane but different heading direction and the conflict exists for CAV~$i\in \mathcal{N}(t)$ only until $\mathcal{D}$. Let $\mathcal{DC}_i(t)$ represent the set of vehicles with higher priority than $i$ arriving in the same lane with different heading direction.  
Merging conflict occurs when vehicles from different entry lanes have the same exit lane. With the knowledge of intersection geometry, entry lane and heading direction from other CAVs, the exit lane can be easily found. A merging conflict is active only in the exit lane starting from $\mathcal{M}$. Let $\mathcal{MC}_i(t)$ represent the set of CAVs with higher priority than $i$ having the same exit lane.
A crossing conflict occurs inside the IZ when CAVs with different paths intersect. From Fig.~\ref{fig:CC_points} there are a total of 16 crossing points, and $\mathcal{CC}_i(t)$ represents a set containing the higher prioritized CAVs whose paths intersect with CAV~$i$.
The sequential/car-following conflict occurs when vehicle $i$ has a preceding CAV~$i$ following the same path. 
$\mathcal{CF}_i(t)$ represents the set of vehicles with higher priority that have the same path as $i$.
The sets $\mathcal{DC}_i(t),\mathcal{MC}_i(t),\mathcal{CC}_i(t),\mathcal{CF}_i(t)$ are ordered with ascending order of priority defined by $\mathcal{CO}(t)$ and are subsets of $\mathcal{N}(t)$. They are collectively referred to as the conflicting sets in the reminder of the paper.

Another important constraint independent of the other CAVs in the intersection is the turning speed constraint. We enforce a safe turning speed for CAVs taking a turn in the intersection. The centripetal force provided by the tyre friction forces, defines the safe speed of the CAVs in turns and is given by, 
\begin{equation}\label{trspeed}
    v_{safe} = \sqrt{fgR} ,
\end{equation}
where $f$, $g$, and $R$ represent the tyre friction coefficient, gravitational constant and radius of the turn. The latter is different for left and right turning vehicles.

It now remains to obtain the optimal feedback control law $u_i(t)$ for each CAV satisfying the above constraints.

\section{Optimal Control Problem Formulation in an Intersection}
This section describes the Optimal Control Problem (OCP) formulation for a CAV~$i$ and the mathematical translation of the various conflicts previously discussed. The solution is obtained using Pontryagin's Minimum Principle (PMP).  

\subsection{Unconstrained Eco-Driving Problem}
With the main objective of ED to minimize cumulative energy consumption over a trip and with our work focusing on EVs, the running cost is given by the electrochemical power from the battery $P_b(t)$ defined in (\ref{Pb}). With only the main equations summarized here, the readers are referred to (\cite{Han.2019,sciarretta2020energy}) for a detailed derivation. The optimal control problem reads,

\begin{multline}\label{OCP}
   a_i(t) = \operatorname*{arg\;min}_{a_i(t)} \int_{0}^{T_i} p_0(a_i(t+k) + h)v_i(t+k) \\ + p_1(a_{i}(t+k) + h)^2\,dk
\end{multline}
subject to  state dynamics, 
\begin{equation*}\label{model}
     \dot{x}_i = v_i(t),\;\;  \dot{v}_i = a_i(t),\;
\end{equation*}
and Boundary Conditions ($\mathcal{BC}$),
\begin{equation*}\label{BCs}
  x_i(t),\;\;  v_i(t),\;\;  x_i(T_i) = D_i,\;\;  v_i(T_i) = V_i, 
\end{equation*}
where $x_i(t)$ and $v_i(t)$ are the position and velocity at time $t$. The desired final time is denoted by $T_i$ and $D_i$ is the length of the path from the start of the entry lane to the end of the exit lane. The desired final speed is $V_i$. 
The solution of (\ref{OCP}) yields a parabolic speed profile as a function of time,
\begin{multline}\label{unconstrained}
    v_i(t+k) = v_i(t) + \left(-\frac{4v_i(t)}{T_i} - \frac{2V_i}{T_i} + \frac{6D_i}{T_i^2}\right)k+ \\
    \left(\frac{3v_i(t)}{T_i^2}- \frac{6D_i}{T_i^3} + \frac{3V_i}{T_i^2}\right)k^2, \;\; k\in[0,T_i)\;.
\end{multline}
and a cubic position profile as a function of time,
\begin{multline}\label{unconstrained_position}
    x_i(t+k) =  x_i(t) + v_i(t)k + \left(-\frac{2v_i(t)}{T_i} - \frac{V_i}{T_i} + \frac{3D_i}{T_i^2}\right)k^2+ \\
    \left(\frac{v_i(t)}{T_i^2}- \frac{2D_i}{T_i^3} + \frac{V_i}{T_i^2}\right)k^3, \;\; k\in[0,T_i)\;.
\end{multline}

The associated energy consumption of the trip $E_b$, a function of vehicle parameters $p_0$, $p_1$, $h$ and boundary conditions, is given by
\begin{equation}\label{Energy unconstrained}
\begin{aligned}
E_{b} &=p_0 h D_{i}+p_0 \frac{V_{i}^{2}-v_{i}^{2}(t)}{2}+p_1 h^{2} T_{i}+2 p_1 h\left(V_{i}-v_{i}(t)\right)+\\
&+4 p_1\left(\frac{3 D_{i}^{2}}{T_{i}^{3}}-\frac{3 D_{i}\left(v_{i}(t)+V_{i}\right)}{T_{i}^{2}}+\frac{v_{i}^{2}(t)+v_{i}(t) V_{i}+V_{i}^{2}}{T_{i}}\right) .
\end{aligned}
\end{equation}

\subsection{Car-Following conflict/Sequential Conflict}\label{CFC}
Car-following conflict occurs when CAV~$i$ and an immediately preceding CAV $p \in \mathcal{CF}_i(t)$, have the same path. Such a conflict leads to a potential rear-end collision without adjusting CAV~$i$’s speed. A collision of such type is formulated as a position-inequality constraint. CAV~$p$’s motion is predicted under the assumption of constant acceleration $a_p(t)$ for the entire horizon $T_i$,
\begin{equation}\label{PC}
x_i(t+k)  \leq  x_{p}(t) + v_{p}(t)k + \frac{1}{2}a_{p}(t)k^2  - s_{min}\;,
\end{equation}
where $x_p(t)$, $v_p(t)$ and $a_p(t)$  are the position, velocity and acceleration of CAV~$p$, while $s_{min}$ denotes the constant safe minimum gap. In the rest of the paper, the term $x_p(t) - s_{min}$ is lumped into a single term $x_p(t)$ for convenience. 
The car-following constraint is solved in (\cite{8286942}) and the main equations are presented here. The optimal speed profile is given as
\begin{multline}\label{constrained}
 v_i(t+k) = 
   v_i(t) + \Big( a_{p}(t) + \frac{4}{\theta_i}(v_p(t) - v_i(t)) +  \\\frac{6}{\theta_i^2}(x_p(t) - x_i(t))\Big)k - 
   \Big( \frac{6}{\theta_i^3}(x_p(t) - x_i(t)) +\\ \frac{3}{\theta_i^2}(v_p(t) - v_i(t)) \Big)k^2,\;\; k\in[0,\theta_i)\;,
\end{multline}
where $\theta_i$ denotes the contact time where the position constraint is met ($x_i(t + \theta_i) = x_p(t)$) and is found by solving the cubic equation 
\begin{multline}\label{t1}
\Big(v_{i}(t) - V_i + a_{p}(t) T_i\Big) \theta_i^{3}+\\
\Big(4 v_{p}(t) T_i+ V_i T_i -2 v_{i}(t) T_i + \frac{a_{p}(t) T_i^{2}}{2}-3 D_i\Big) \theta_i^{2}+\\
\Big(6 (x_p(t) - x_i(t)) T_i + v_{i}(t) T_i^{2}-v_{p}(t) T_i^{2}\Big) \theta_i- \\ 3(x_p(t) - x_i(t)) T_i^{2}=0.
\end{multline}

The associated closed-form energy consumption is not presented here due to space constraints but is easily obtained by inserting (\ref{constrained}) in (\ref{OCP}).

\subsection{Crossing Conflict}
A crossing conflict occurs when the path of two or more CAVs intersects in the IZ leading to a potential lateral collision between them. A collision of such type is formulated as an interior-point constraint (\cite{BrysonHo69}), where equality constraints on the states are imposed point-wise along the horizon.  Consider two CAVs $i$ and $c \in \mathcal{CC}_i(t)$ with $\mathcal{CO}(t) = \{c,i\}$. The lateral collision avoidance constraint is formulated as
 \begin{equation}\label{IPC}
        x_i(t + K_c^{\mathcal{C}_{ci}}) = \mathcal{C}_{ic} - x_i(t) \;\;\;  \text{with}  \;\; K_c^{\mathcal{C}_{ci}} = k_c^{\mathcal{C}_{ci}} + dT\;,
 \end{equation}
where $\mathcal{C}_{ic}$ is the distance to the crossing point from $x_i(0)=0$, $k_c^{\mathcal{C}_{ci}}$ is the arrival time of CAV $c$ at its own crossing point $\mathcal{C}_{ci}$, and $dT$ is a safety margin between arrival of the CAVs at their respective crossing points.
The solution to the above constraint is obtained as follows. The optimal speed profile of CAV~$i$ is composed of two almost-independent unconstrained segments (\ref{unconstrained}), defined by $\mathcal{BC}$: $x_i(t),v_i(t), \mathcal{C}_{ic},v_i(t + K_c^{\mathcal{C}_{ci}})$ for the first segment and $\mathcal{C}_{ic}, v_i(t + K_c^{\mathcal{C}_{ci}}), D_i,,V_i$ for the second segment.  With the position and time at the junction of the two segments imposed by (\ref{IPC}), the optimal speed profile is completely defined by the free parameter $v_i(t + K_c^{\mathcal{C}_{ci}})$. The energy consumption of the whole trip $E_b$ is the sum of energy consumption of each segment and the optimal value for $v_{it} = v_i(t + K_c^{\mathcal{C}_{ci}})$ is obtained by minimizing $E_b$, 
 \begin{equation}\label{ccEb}
 \begin{aligned}
     E_b &= E_{b}^{(1)} + E_{b}^{(2)}, \\
     v_{it}\; &\text{s.t.}\; \partial E_b/\partial v_{it} = 0
 \end{aligned}
 \end{equation}
where $E_{b}^{(1)}$ and $E_{b}^{(2)}$ are obtained from (\ref{Energy unconstrained}). $v_{it}$ is explicitly obtained and is a function of the two segments' boundary conditions.The optimality of the approach has been shown in (\cite{Han.2019,sciarretta2020energy}).    

\subsection{Diverging Conflict}\label{DC}

\begin{figure}[h]
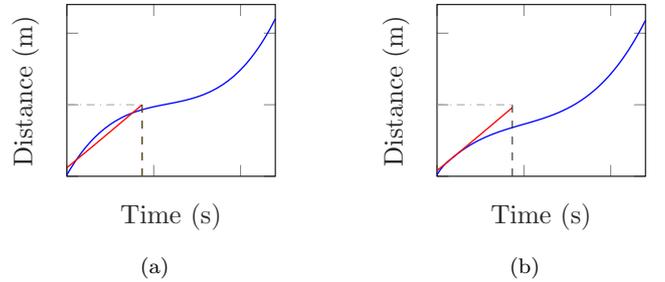

     \begin{subfigure}[h]{0.45\columnwidth}
         \input{images/cuu_constrained}
         \caption{}
         \label{fig:DP conflict cons}
     \end{subfigure}
     ~\hfill
     \begin{subfigure}[h]{0.45\columnwidth}
         \input{images/cuu_solved}
         \caption{}
         \label{fig:DP  conflict sol}
     \end{subfigure}
     \caption{Diverging Conflict Solution. Fig.\ref{fig:DP conflict cons}, shows constraint violation when employing (\ref{unconstrained})and in fig. \ref{fig:DP conflict cons}, CAV~$i$ uses the solution obtained from (\ref{Diveringsol}). CAV~$d$ and CAV~$i$ are represented by red and blue respectively}
     \label{cuu}
\end{figure}

A diverging conflict occurs, in the entry lane, when CAV~$i$ and its immediately preceding vehicle $d \in \mathcal{DC}_i(t)$ have the same entry lane and different heading direction. Once CAV~$d$ changes direction at $\mathcal{D}$, CAV~$i$ is no longer in conflict. Like the car-following, the diverging conflict poses a potential rear-end collision and is formulated as a position-inequality constraint with the exception that CAV~$i$ is constrained only until $\mathcal{D}$, see Fig.~\ref{cuu}. More formally,
 
\begin{equation}\label{divercon}
   x_i(t+k) \leq \left\{
        \begin{array}{ll}
            x_{d}(t) + v_{d}(t)k + \frac{1}{2}a_{d}(t)k^2  - s_{min} & \quad x_i(t) \leq \mathcal{D} \\
            \infty & \quad x_i(t) >  \mathcal{D}
        \end{array}
    \right.
\end{equation} 
 As in section \ref{CFC}, CAV~$d$’s motion is predicted under the assumption of a constant acceleration until $\mathcal{D}$. 
 
 The optimal solution to the above constrained OCP is made up of two segments, i.e., a first segment where (\ref{constrained}) applies, satisfying the inequality constraint and the second segment where (\ref{unconstrained}) applies.  The $\mathcal{BC}$ for the first and second arc are given as $x_i(t), v_i(t),x_i(t + k_d^{\mathcal{D}}),v_i(t + k_d^{\mathcal{D}})$ and $x_i(t + k_d^{\mathcal{D}}),v_i(t + k_d^{\mathcal{D}}),D_i,V_i$ respectively. With the time at the junction of the segments imposed by $k_d^{\mathcal{D}}$ (arrival time of CAV~$d$ at $\mathcal{D}$), the optimal solution is now defined by two unknown free parameters, $x_i(t + k_p^{\mathcal{D}}),v_i(t + k_p^{\mathcal{D}})$. The optimal value for these two parameters is found by minimizing the sum of energies of the two segments (i.e., the total energy consumption of the trip $E_b$).
 \begin{equation}\label{Diveringsol}
 \begin{aligned}
     E_b &= E_{b}^{(1)} + E_{b}^{(2)}, \\
    v_i(t+ k_d^{\mathcal{D}}) \; &\text{s.t.}\; \partial E_b/\partial v_i(t+ k_d^{\mathcal{D}}) = 0,\\
    x_i(t+ k_d^{\mathcal{D}}) \; &\text{s.t.}\; \partial E_b/\partial x_i(t+ k_d^{\mathcal{D}}) = 0
 \end{aligned}
 \end{equation}
 where $E_{b}^{(1)}$ is energy consumption obtained using the solution (\ref{constrained}) and $E_{b}^{(2)}$ is given by (\ref{Energy unconstrained}).

\subsection{Merging Conflict}
A merging conflict occurs when the path of CAV~$i$ and CAV $e \in \mathcal{MC}_i(t)$, merge at $\mathcal{M}_i$. 
CAV~$i$ remains conflict free from CAV~$e$ until $\mathcal{M}_i$ but upon entering the exit lane CAV~$e$ poses a potential rear-end collision. As in section \ref{DC}, merging conflict is formulated as a position- inequality constraint but starting from $\mathcal{M}_i$ until the end of the horizon.The subscript $i$ in $\mathcal{M}_i$ indicates that the merging point depends on the path (i.e., left, right and straight) taken by the CAV~$i$.

The optimal solution to the above constrained OCP is made up of two segments, i.e., first segment where (\ref{unconstrained}) applies and the second segment where (\ref{constrained}) applies. The $\mathcal{BC}$ for the first and second arc are given as $x_i(t), v_i(t), \mathcal{M}_i,v_i(t + k_e)$ and $\mathcal{M}_i,v_i(t + k_e),D_i,V_i$, respectively. CAV~$i$'s position at the end of first segment is constrained to arrive at $x_i(t + k_e) = \mathcal{M}_i$, where $k_e$ is the arrival time of CAV~$e$ at $\mathcal{M}_e + ds \; \text{s.t.} \; ds > s_{min}$. This ensures that CAV~$i$ arrives at $\mathcal{M}_i$ after $e$ and does not violate the position constraint at $\mathcal{M}_i$. With both position and  time at the junction imposed, the optimal solution is now defined by the one unknown free parameter, $v_i(t + k_e)$. The optimal value for this parameter is found by minimizing the trip's total energy consumption.

 \begin{equation}
 \begin{aligned}
     E_b &= E_{b}^{(1)} + E_{b}^{(2)}, \\
     v_{i}(t+k_e)\; &\text{s.t.}\; \partial E_b/\partial v_{i}(t+k_e) = 0
 \end{aligned}
 \end{equation}

where, $E_{b}^{(2)}$ is the energy consumption obtained using the solution (\ref{constrained}) and $E_{b}^{(1)}$ is given by (\ref{Energy unconstrained}). 

\begin{rem}
A CAV $i \in \mathcal{N}(t)$, could face none, any one or a combination of the four basic conflicts. When there is a combination of conflicts, the optimal solution consists of two or more segments. The free parameters at the junction are identified and the optimal value for the free parameters are obtained by minimizing the energy of the entire trip, as shown earlier. An exhaustive list of all possible combinations and its solutions are not presented here due to limitations in space.    
\end{rem}

\subsection{Turning Constraint}
The turning speed constraint restricts a CAV's speed in the IZ to a maximum defined by (\ref{trspeed}). The constraint is enforced by letting the speed of CAV~$i$ equal to $v_{tr} < v_{safe}$ at the mid point of its path in the IZ. The choice of $v_{tr}$, a lower value than $v_{safe}$ is a conservative approach that keeps the CAV's speed below the threshold. More formally,
\begin{equation}
\begin{aligned}
    v_i(t+k_t) &= v_{tr},\\
    x_i(t+k_t) &= (D_i + x_i(t)) /2,
\end{aligned}
\end{equation}
and $v_{tr}$ is obtained from
\begin{equation}
    \frac{\delta}{2}  - \left(\frac{v_{tr}(v_{safe} - v_{tr})}{a_{max}} + \frac{(v_{safe} - v_{tr})^2}{2a_{max}}\right) = 0\;,
\end{equation}
 where $a_{max}$ is the maximum acceleration of CAV~$i$ and $\delta$ is the vehicle's turning path inside the IZ.  The free parameter $k_t$ is obtained by replacing  $v_{it}$ by $k_t$ by in (\ref{ccEb}), with $\mathcal{BC}$ for the two segments given by $x_i(t),v_i(t),(D_i + x_i(t)) /2, v_{tr}$ and $(D_i + x_i(t)) /2, v_{tr}, D_i,,V_i$.

\section{Algorithm}
This section describes how, when a CAV~$i$ enters the intersection, it identifies the constraints posed by the other CAVs already in the intersection, and obtains its optimal solution. 
At each time $t$, vehicles in the intersection, i.e., $\mathcal{N}(t) $, solve their OCP sequentially in the order dictated by $\mathcal{CO}(t)$ . Therefore, when CAV $i \in \{ \mathcal{N}(t) $ : $|\mathcal{N}(t) > 1 | \}$ enters the intersection, all higher prioritized vehicles have already solved their OCP and hence can share their control inputs to other arriving vehicles if needed. 
CAV~$i$, using the lane and heading direction information shared by other vehicles and $\mathcal{CO}(t)$ from the coordinator, can compute the vehicles in the conflicting sets.  In the absence of any vehicle in the intersection or any vehicle belonging to the conflicting sets, CAV~$i$ is unconstrained and can simply follow the solution in (\ref{unconstrained}). However, in the presence of one or more CAVs in the conflicting sets, CAV~$i$ could either face one of the four basic constraints or a combination of them. 
If there is only one vehicle that belongs to any one of the conflicting sets, CAV~$i$ has one of the four basic constraints. On the other hand, if there is one vehicle from more than one of the conflicting sets, then CAV~$i$ has a combination of the four constraints. For example, CAV $ j \in \mathcal{DC}_i(t) $ and CAV  $ k \in \mathcal{LC}_i(t)$, means CAV~$i$ has a position-inequality constraint until $\mathcal{D}$ and an interior-point constraint in the IZ. All possible combinations of conflicts are identified, \textit{a priori}, and their solutions are computed as described in the previous section. 
With the conflicts posed by the higher priority vehicles identified, all possible solutions satisfying the conflicts are evaluated and CAV~$i$ applies the solution with the least energy consumption.
Using the above example, the possible solutions to the constraint includes the unconstrained solution (\ref{unconstrained}), the solution avoiding the diverging conflict, the solution avoiding the crossing conflict, and the solution avoiding both. All the four possible solutions are evaluated to choose the one consuming the least energy without violating any constraint. The solution to these conflicts require information from the corresponding CAVs to predict their motion or arrival time. The amount of information received depends on the cooperation amongst the CAVs
\begin{rem}
If $|\mathcal{DC}_i(t) \cup \mathcal{LC}_i(t)| > 1$, then CAV~$i$ is in conflict only with its immediately preceding vehicle, i.e., CAV with the lowest priority in $|\mathcal{DC}_i(t) \cup \mathcal{LC}_i(t)|$. It should be noted that CAV~$i$ can either face a diverging conflict or a car-following conflict and not both simultaneously.  If $|\mathcal{MC}_i(t)| > 1$, then CAV~$i$ is in merging conflict with the last merging vehicle, i.e. CAV with lowest priority in $|\mathcal{MC}_i(t)|$. If $|\mathcal{LC}_i(t)| > 1$, the CAV that causes the maximum violation of the unconstrained solution of CAV~$i$, a method inspired from (\cite{VANKEULEN2014187}), is the conflicting CAV. 
\end{rem}

The described algorithm is embodied in a shrinking horizon MPC fashion, where the boundary conditions are updated at each time step. 

We consider two levels of cooperation, namely the Non-Cooperative ED (NC-ED) and the Cooperative ED (C-ED) scenario. The scenarios differs in two aspects : (\romannum{1}) the point at which CAV~$i$ determines if a higher prioritized vehicle poses a crossing or merging conflict and (\romannum{2}) the amount of information shared to CAV~$i$. As mentioned earlier, CAVs in the intersection solve their OCP sequentially in the order dictated by $\mathcal{CO}(t)$. Therefore, when CAV~$i$ solves for its OCP, all higher prioritized CAVs all have their solutions. Based on CAV~$i$'s active constraint, it might either need the conflicting CAV's motion to avoid a rear-end collision or its arrival time at a crossing point to avoid a lateral collision or both. 

In the NC-ED scenario, CAV~$i$ decelerates to a fixed distance before the $\mathcal{D}$ called the visibility distance $D_{vis}$, at which it determines whether a higher prioritized CAV poses a crossing or merging conflict. This approach is followed in (\cite{sumo}). The speed at $D_{vis}$ is equal to $a_{max}$ multiplied by one second.   Up until $D_{vis}$, in the presence of either a diverging conflict or a car-following conflict, CAV~$i$ measures the instantaneous acceleration of CAV~$d$ or $p$ to predict its motion as in (\ref{divercon}) or (\ref{PC}). At $D_{vis}$, CAV~$i$, in the presence of a merging conflict, measures the instantaneous acceleration $a_e(t)$ of CAV $e \in \mathcal{MC}_i(t)$. The acceleration $a_e(t)$, is used to predict CAV~$e$'s arrival time $k_e$ and its  motion in the exit lane. In the presence of a crossing conflict, CAV~$i$ measures the instantaneous acceleration $a_c(t)$ of CAV $c \in \mathcal{CC}_i(t)$, to predict its arrival time  $k_c^{\mathcal{C}_{ci}}$. The motion and arrival time of the conflicting vehicles are predicted under a \textit{constant acceleration} assumption in the NC-ED.

In the C-ED, CAV~$i$ determines if the higher prioritized vehicles present a conflict or not as soon as it enters the intersection. Cooperation is introduced in amongst CAVs as the ability to share the intentions over a certain horizon. In the presence of a conflict causing a rear-end collision, the CAVs share their solution to CAV~$i$. In order to be used in the eco-driving control of CAV~$i$, this information is lumped into one ``future mean value" $\tilde{a}_{x}$, evaluated over a preview window length $L$, as
\begin{equation}
\tilde{a}_{x}(t) = \frac{1}{L} \int_{t}^{t+L} a_{x}(\tau)d\tau\;\; x \in \{d,p,e\} .
\end{equation}
as in (\cite{9564680}). The acceleration $\tilde{a}_{x}(t)$, replaces the measured instantaneous acceleration in the non-cooperative scenario, to predict the conflicting CAV's motion.
The arrival times $k_c^{\mathcal{C}_{ci}}$ and $k_e$ are computed from the actual arrival time given by the solution used by CAVs $c$ and $e$. For example, if CAV $c$'s solution is given by (\ref{unconstrained}), its arrival time at $\mathcal{C}_{ci}$ is computed using (\ref{unconstrained_position}). 

\section{Simulation}
In this section, we evaluate the performance of the proposed NC-ED and C-ED algorithm to cross an urban intersection, in a MATLAB simulation environment. The algorithms are compared to a baseline scenario generated by the traffic simulation software SUMO. The Intelligent Driver Model (IDM) is used to describe the behavior of the CAVs in the baseline.  The performance is evaluated, in terms of energy consumption, using a detailed Nissan Leaf electric vehicle model presented in (\cite{dollar}).

We consider an isolated urban intersection where \textit{l} = 47~m and \textit{w} = 4~m. The minimum safety distance $s_{min}$ = 7~m and the safety $dT$ to avoid a lateral collision is 2.5~s. 
The initial and final speeds are uniformly distributed in the interval 8.3~m/s $\pm$ 2~m/s. The maximum acceleration and deceleration of the CAVs are set to 4~m/s$^2$. A total of 30 CAVs are simulated with a traffic flow rate of 800, 1200, and 1600~veh/hr. The entry time of the CAVs follows an exponential distribution and the lanes and heading directions are assigned using a uniform distribution. Unlike the ED solutions, the IDM cannot enforce a prescribed final time. The final time of CAV~$i$ obtained as an output of IDM is used in ED to ensure same average speed for a CAV~$i$ across the algorithms. The crossing order is given by the right-before-left criterion in SUMO. 
\begin{figure}[ht]
    \centering
%
%
\definecolor{mycolor1}{rgb}{0.00000,0.44700,0.74100}%
\definecolor{mycolor2}{rgb}{0.85000,0.32500,0.09800}%
\definecolor{mycolor3}{rgb}{0.92900,0.69400,0.12500}%
\begin{tikzpicture}

\begin{axis}[%
width=4.5in,
height=3.5in,
at={(0.758in,0.481in)},
scale = 0.6,
xmin=0.5,
xmax=3.5,
xtick={1,2,3},
xticklabels={{800},{1200 },{1600 }},
xlabel style={font=\color{white!15!black}},
xlabel={Flow Rate (veh/hr)},
ylabel style={font=\color{white!15!black}},
ylabel={Total Energy (kJ)},
ymin=1.300,
ymax=2.000,
axis background/.style={fill=white},
xmajorgrids,
ymajorgrids,
legend style={at={(axis cs:0.53,1.99)},anchor=north west, legend columns =3, nodes = {scale=0.65}}
]
\addplot [color=mycolor1, only marks, mark size=2.0pt, mark=*, mark options={solid, mycolor1}]
  table[row sep=crcr]{%
1	1.7719\\
2	1.8693\\
3	1.921\\
};
\addlegendentry{IDM}

\addplot [color=mycolor2, only marks, mark size=2.0pt, mark=*, mark options={solid, fill=mycolor2, mycolor2}]
  table[row sep=crcr]{%
1	1.7174\\
2	1.73386\\
3	1.8409\\
};
\addlegendentry{NC-ED}

\addplot [color=mycolor3, only marks, mark size=2.0pt, mark=*, mark options={solid, fill=mycolor3, mycolor3}]
  table[row sep=crcr]{%
1	1.3504\\
2	1.42143\\
3	1.4708\\
};
\addlegendentry{C-ED}
\end{axis}
\end{tikzpicture}%
    \caption{Energy Consumption}
    \label{fig:res}
\end{figure}
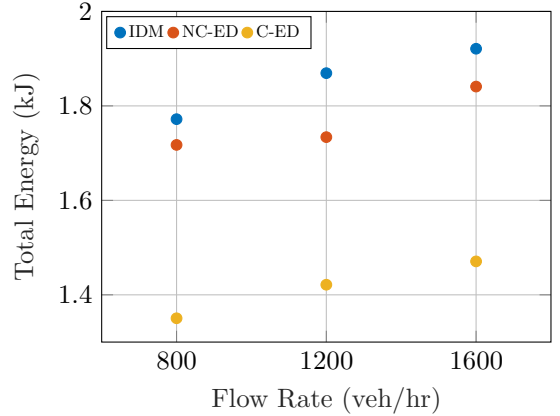

\begin{figure}[h]
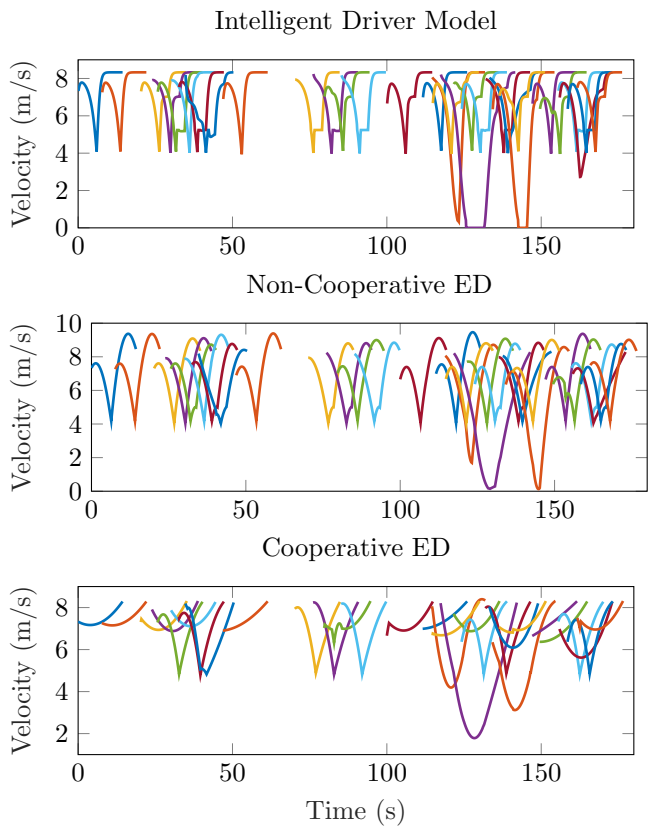

    \centering
    \begin{subfigure}{\columnwidth}
    \input{images/IDM}
    \label{fig:IDM}
    \end{subfigure}
    \begin{subfigure}{\columnwidth}
    \input{images/noncoop_800}
    \label{fig:noncoop}
    \end{subfigure}
    \begin{subfigure}{\columnwidth}
    \input{images/coop_800}
    \label{fig:coop}
    \end{subfigure}
\caption{Velocity trajectories of CAVs at 800~veh/hr}
    \label{fig:velocity}
\end{figure}

The velocity trajectories of the CAVs, arriving at a flow rate of 800~veh/hr, equipped with IDM, NC-ED and C-ED algorithms are shown in Fig.~\ref{fig:velocity}. The CAVs using IDM and NC-ED decelerate to 4~m/s$^2$ at $D_{vis}$, which is 4.5~m before $\mathcal{D}$. At $D_{vis}$, CAV~$i$  checks for merging or crossing conflicts with higher prioritized CAVs and obtains only the instantaneous control input from the conflicting CAVs.  In the event of a conflict, the CAV~$i$ adjust its speed profile, with some of them coming to an almost complete stop, e.g. the purple CAV at 130~s. On the other hand CAVs with C-ED have less variations in acceleration. CAVs taking left or right turns also decelerate to satisfy turning speed constraints, e.g. CAV~5, green speed profile at around 40~s, is taking a right turn and has a safe speed of 5.24~m/s inside the IZ. 

Each flow rate is simulated thrice with randomly generated boundary conditions and their energy consumption are given in Fig.~\ref{fig:res}. It can been seen that the total energy consumption increases with an increase in flow rate. This is due to the increased conflicts amongst the CAVs causing a CAV~$i$ to adjust its speed profile.
The CAVs in NC-ED consumes 3~$\%$ less energy than IDM.  With IDM and NC-ED having same information on conflicting vehicles, the energy reduction is caused by the optimal ED speed profiles employed by the CAVs. The C-ED performs best amongst the three, with a reduction of 23.7~$\%$ over IDM and 21.3~$\%$ over the NC-ED. The CAVs with cooperation have more accurate information about the conflicting vehicles and get this information much earlier (i.e. at $x_i(0) =0$) than the other two algorithms. This enables better anticipation and hence the CAVs can adjust their speed profiles accordingly. 


\section{Conclusion}
The paper addressed the problem of finding the optimal ED speed trajectory for the entire horizon (i.e., entry lane, IZ and exit lane) of an urban isolated intersection. Various conflicts in the intersection were formulated has as constraints to the CAV and its solutions were presented. Two levels of cooperation, namely, the non-cooperative and cooperative, were studied and evaluated for  energy consumption in comparison to the baseline IDM. The cooperative algorithm performed best indicating that, earlier determination of conflicts with with higher prioritized CAVs and better prediction of their arrival time and motion, leads to better energy efficiency.

\bibliography{intersection_biblio}
\end{document}